\documentclass[twoside]{article}
\usepackage{amsmath, amsfonts, amssymb}
\usepackage{epsf}
\usepackage{graphicx}
\usepackage{aat}
\usepackage{tabularx}
\usepackage{aattable}
\renewcommand{\deg}{$^\circ$}

\begin{document}

\thispagestyle{myfirst}

\setcounter{page}{1}

\mytitle{Properties and Structural Features of Early-type Disk Galaxies 
with Multi-tier Disks} \myauthor{M.A. Ilyina, O.K.Sil'chenko}
\myadress{Sternberg Astronomical
 Institute, 13, Universitetskij pr., Moscow 119991, Russia}
\mydate{(Received October , 2010)}
\myabstract
{The results of photometric decomposition of 85 early-type galaxies
are presented. The SDSS $r$-images are analysed. Double-tier exponential disks 
are found in all galaxies which are studied; the statistics of disk parameters 
is analysed.
}
\mykey{galaxies: structure, evolution}

\section{Introduction}

The photometric decomposition of disk galaxy images is a long-standing 
classical topic in studies of galaxy structure. However, the Sloan Digital
Sky Survey (SDSS) has provided new possibilities due to its depth 
and large sky area covered by the observations. The statistics of
structure characteristics obatined with the SDSS data for nearby late-type 
galaxies has been already analysed \cite{pt2006}. Here we present some 
results for a sample of early-type disk galaxies.

\section{Sample}

For our decomposition we have selected 85 galaxies satisfying the
following criteria:

\begin{itemize}

\item{
the morphological type of S0--Sa, to avoid isophote distortion 
due to dust;
}
\item{
disk inclination less than 70\deg, to avoid degeneracy owing to disk light
integration at our line of sight;
}
\item{
redshift less than 0.03, to provide sifficient metric spatial resolution;
}
\item{
the absence of large-scale bar (though we do not reject the galaxies
with nuclear minibars which do not affect disk parameter determinations).}

\end{itemize}

For these 85 galaxies, the DR7 data of the SDSS imaging \cite{sdssdr7} 
have been retrieved. We have chosen $r$-images because they have the best 
signal-to-noise ratio for our reddish galaxies. The software used for 
decomposition is GIDRA (Galaxies
Interactive DRAwing) which allows to build iteratively one-dimensional surface
brightness profiles by averaging counts per pixel over elliptical rings and
to construct 2D model maps described by these profiles to subtract them from
the images observed \cite{moiseev}. A combination of components tried 
to fit every galaxy includes a Sersic bulge and two exponential disks with 
different scalelengths.

\section{Results}

The most interesting result of the decomposition of 85 early-type disk galaxies
is that all of them possess double-tier large-scale disks; it is seen in the
band $r$ best of all while in the band $g$ the outer disks with a larger scalelength
are sometimes lost among statistical noises due to low surface brightness.

\subsection{Classification}

\begin{figure}
\resizebox{\hsize}{!}{\includegraphics{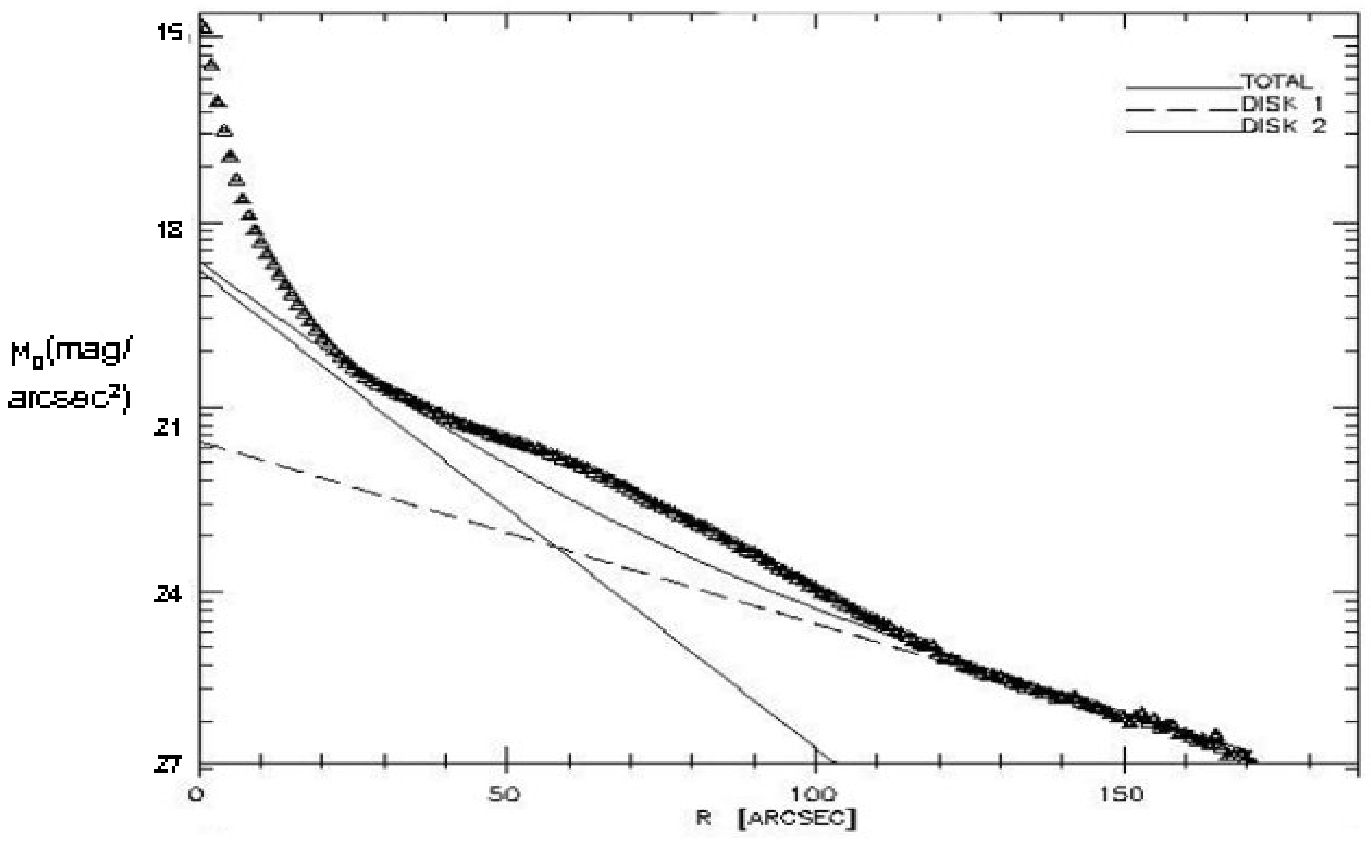} 
\includegraphics{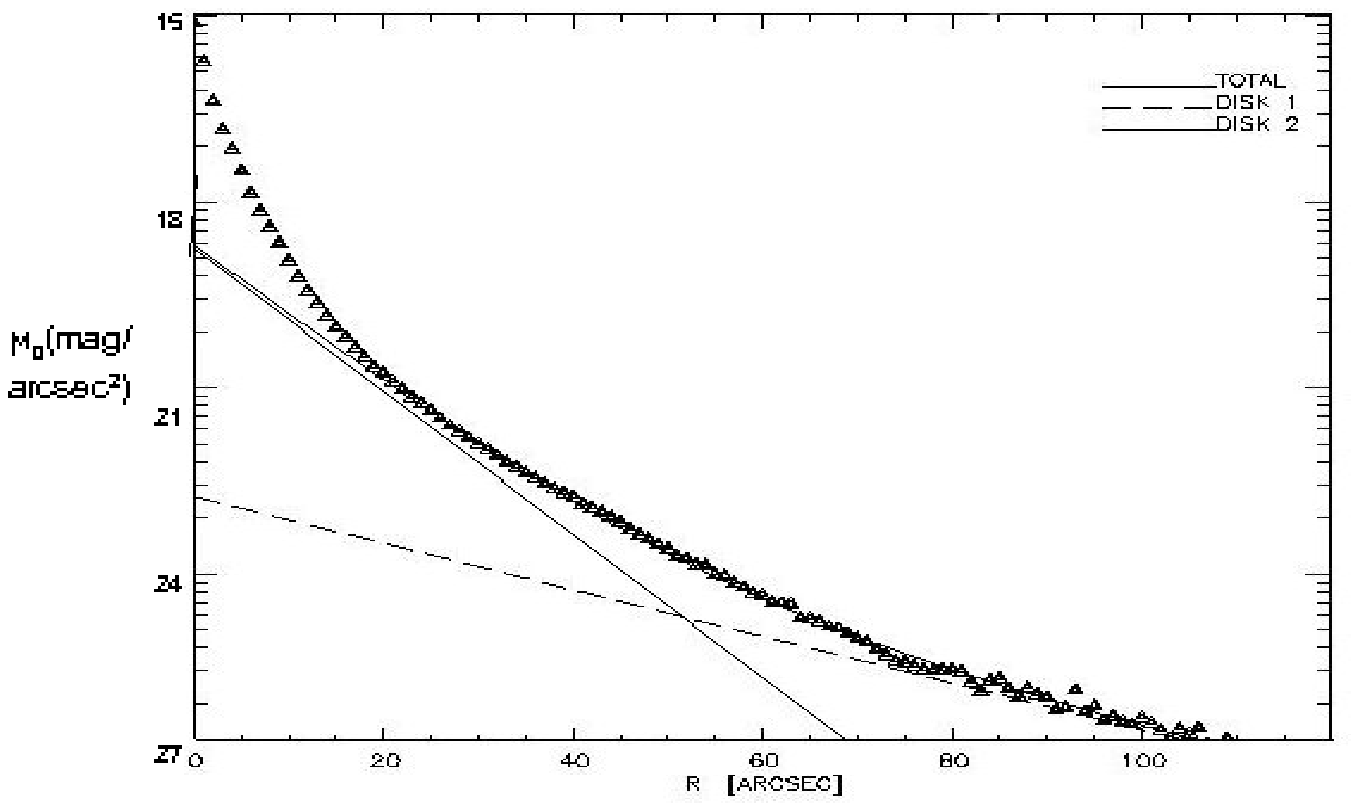}}
\caption{Typical surface brightness profiles in galaxies of our sample:
{\it left} -- NGC~4203, so called `truncated' inner disk which we treat as an 
exponential disk containing stellar ring; {\it right} -- NGC~4169, 
so called `antitruncated', or double-tier, disk fitted by two exponential segments 
with different scalelength, the outer one being longer.}
\end{figure}

By following the classification scheme by Erwin et al. \cite{erwin}, we try to
divide all the profiles obtained into 3 groups:

\begin{itemize}

\item{classical one-scale exponential law; we have no such galaxies
in our sample;
}
\item{truncated inner disks; they demonstrate a shallow cut-off at some radii;
but in our collection of the deep SDSS $r$-images the truncated inner disks are
always followed by outer disks with a larger scalelength (Fig. 1a); they
constitute about 32\%\ of all galaxies;  
}
\item{antitruncated disks which are a combination of the inner and outer disks
with different scalelengths, the outer disks having larger scalelengths (Fig. 1b);
they are a majority, about 68\%\ of all sample galaxies.
}

\end{itemize}

Basing on the approach by Sil'chenko \& Moiseev \cite{nucrings}, applied in their paper
to the photometric decomposition of the nuclear-ring galaxy NGC~7742, we have 
suggested that the `truncated' inner disks may be in fact exponential disks 
with the ring-like brightness excess at some radius inside them; a close inspection 
of the images decomposed has confirmed the validity of such approach.

\subsection{Statistics of the disk parameters}

\begin{figure}
\resizebox{\hsize}{!}{\includegraphics{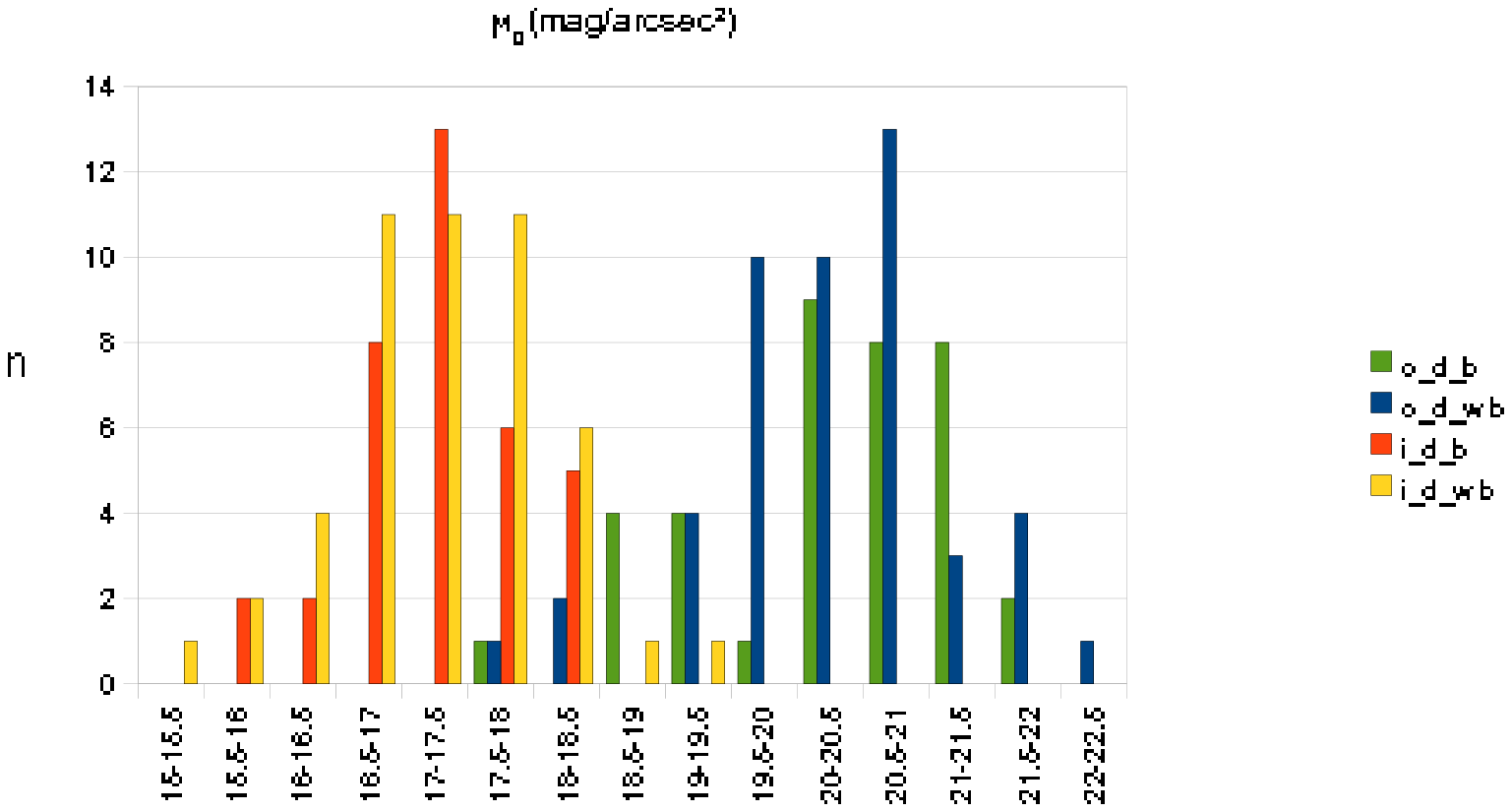} 
\includegraphics{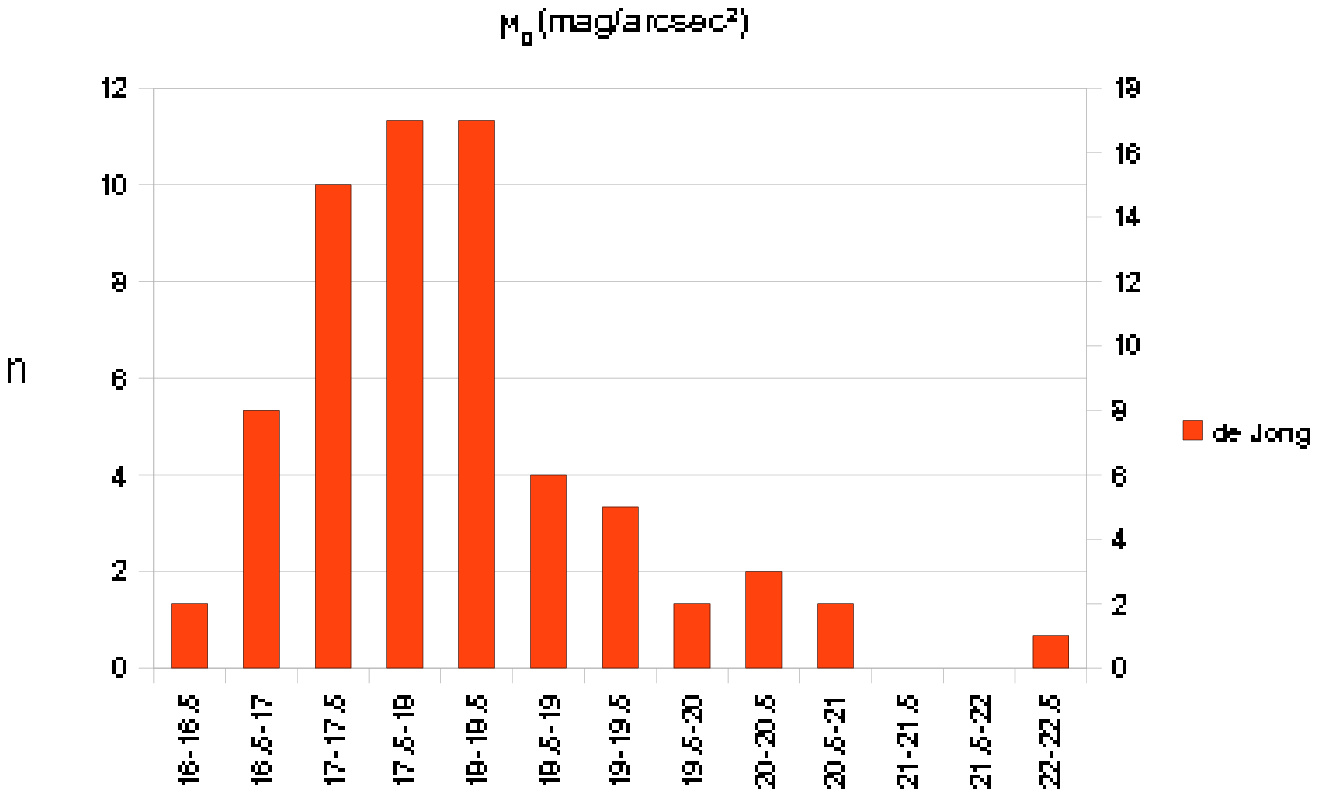}}
\caption{Central surface brightness distributions: {\it left} -- our results,
red colour marks inner disks with nuclear bars, yellow colour marks inner disks without nuclear bars, green colour marks outer disks in galaxies with nuclear bars,
blue colour marks outer disks in galaxies without nuclear bars; {\it right} --
the results from de Jong \cite{dejongdata}. }
\end{figure}

\begin{enumerate}

\item{Figure~2a presents the distribution of inner and outer disks over the
central surface brightnesses which are determined in this work by the decomposition
procedure. The central brightnesses are corrected for the
disk inclinations; for the purpose of comparison with the data of de Jong 
\cite{dejongdata},
the histograms are shifted to the left by about 3~mag to mimic the $K$-band 
measurements (see the SDSS colour transformation equations, \cite{sdsscalib,sdsssyst}).
The distribution is strongly bimodal, the inner-disk
and outer-disk distributions representing almost non-overlapping, Gaussian-like
peaks. Figure~2b shows the similar distribution of the central surface brightnesses
for 86 face-on disk galaxies from \cite{dejongdata}. De Jong \cite{dejongdata} 
fitted 2d images
of the galaxies by a combination of only {\it two} components, both exponential ones,
called `a bulge' and `a disk'. From the comparison of the
histograms in Fig.~2 one can see that the `disk' component in \cite{dejongdata}
corresponds to our inner disks. Obviously, the outer disks of the double-tier 
(`antitruncated') disk galaxies mostly avoided detection up to recent times due to
insufficient depth of the galactic images. Only when the SDSS provided accurate
photometry at the level of 27-28 $r$-mag per square arcsec, the outer disk
detections became quite certain.
}
\item{We have considered separately inner and outer disks in galaxies with nuclear
bars (43\%\ of the sample) and without those (57\%\ of the sample). As Fig.~2a
demonstrates, the distributions of both subsamples over the central surface
brightnesses appear to be similar. 
}
\item{In Fig.~3 we check correlation between the central surface brightnesses
and exponential scalelengths over a totality of large-scale disks of our sample. 
Earlier, many authors who probed photometric decomposition of disk galaxies with 
a single exponential disk found such correlation, in the sense that the more
extended disks had lower surface brightness \cite{dejongstat,iodice,graham,donofrio}.
 However, we noted \cite{n80gr} that this correlation may be an artifact of messing 
inner and outer disks in one data set. Indeed, the outer disks having larger 
scalelengths are mostly low surface brightness disks, as we have seen in Fig.~2a. 
But within two subsamples, this of inner disks and that of outer ones, there is 
no any correlation (Fig.~3). Again, at this diagram we cannot distinguish the 
galaxies with nuclear bars and without ones.
}

\end{enumerate}

\begin{figure}
\centering
\includegraphics[width=6cm]{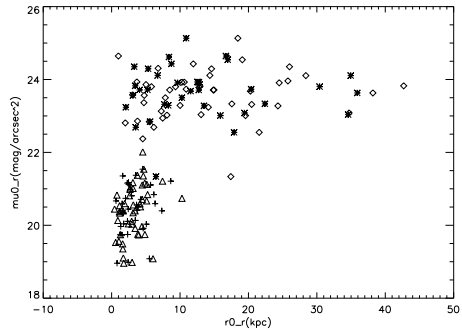} 
\caption{The `central surface brightness--exponential scalelength' diagram
for our galaxies: pluses mark inner disks with nuclear bars, triangles mark inner disks
 without nuclear bars, stars mark outer disks in galaxies with nuclear bars,
diamonds mark outer disks in galaxies without nuclear bars}
\end{figure}

\section{Conclusions}

After having decomposed the SDSS photometric images of 85 early-type disk
galaxies lacking large-scale bars, we conclude:

\begin{itemize}
\item{all early-type disk galaxies in our sample lacking large-scale bars demonstrate
double-tier large-scale stellar disks;
}
\item{the outer and inner stellar disks have probably different nature: the central
surface brightness distribution is strongly bimodal, with the inner and outer
disks having well separated peaks, and at the diagram `$\mu _0$ vs $r_0$'
the outer and inner disks represent two quite separate point clouds;
}
\item{the galaxies with nuclear bars and those lacking them do not separate
at any diagram analyzed here; we conclude that the presence of nuclear bars
does not relate with any structure re-building at large scales.
}

\end{itemize}

The work is supported by the grant of Russian Foundation for Basic Researches
no. 10-02-00062a.



\begin{thebibliography}{}

\bibitem[1]{pt2006}
M. Pohlen and I. Trujillo,  
A\&A {\bf 454} 759 (2006)

\bibitem[2]{sdssdr7}
K.N. Abazajian, J.K. Adelman-McCarthy, M.A. Agueros, {\it et al.},
ApJ Suppl Ser {\bf 182} 543 (2009)

\bibitem[3]{moiseev}
A.V. Moiseev, J.R. Vald\'es, and V.H. Chavushyan,
A\&A {\bf 421} 433 (2004)

\bibitem[4]{erwin}
P. Erwin, M. Pohlen, and J.E. Beckman,
AJ {\bf 135} 20 (2008)

\bibitem[5]{nucrings}
O.K. Sil'chenko and A.V. Moiseev,
AJ {\bf 131} 1336 (2006)

\bibitem[6]{dejongdata}
R.S. de Jong,
A\&A Suppl {\bf 118} 557 (1996a)

\bibitem[7]{sdsscalib}
S. Bilir, S. Ak, S. Karaali, {\it et al.},
MNRAS {\bf 384} 1178 (2008)

\bibitem[8]{sdsssyst}
M. Fukugita, T. Ichikawa, J.E. Gunn, {\it et al.},
AJ {\bf 111} 1748 (1996)

\bibitem[9]{dejongstat}
R.S. de Jong,
A\&A {\bf 313} 45 (1996b)

\bibitem[10]{iodice} 
E. Iodice, M. D'Onofrio, and M. Capaccioli,
Astrophys. Space Sci. {\bf 276} 869 (2001)

\bibitem[11]{graham}
A. Graham and W.J.G. de Blok,
ApJ {\bf 556} 177 (2001)

\bibitem[12]{donofrio}
M. D'Onofrio,
MNRAS {\bf 326} 1517 (2001)

\bibitem[13]{n80gr}
M.A. Startseva, O.K. Sil'chenko, and A.V. Moiseev,
Astronomy Reports {\bf 53} 1101 (2009)

\end{thebibliography}
\end {document}